\documentclass[conference]{IEEEtran}
\pdfoutput=1
\usepackage{graphicx}
\usepackage{makecell}
\usepackage{array}
\usepackage{booktabs}
\usepackage[
    bookmarks=false,
    colorlinks=false,
    hidelinks,
    draft
]{hyperref}
\usepackage{graphicx}
\usepackage{makecell}
\usepackage{comment}

\usepackage{tikz}
\newcommand{\blackcircled}[1]{%
  \tikz[baseline=(char.base)]{
    \node[shape=circle, fill=black, text=white, 
          inner sep=0.1pt,           
          minimum size=0.5em,        
          font=\scriptsize] (char) {#1};  
  }
}

\usepackage{enumitem}
\usepackage{pifont} 

 \usepackage{tcolorbox}
\usepackage{xcolor}
\usepackage{mdframed}

\title{\textit{We Urgently Need Privilege Management in MCP}:
A Measurement of API Usage in MCP Ecosystems}

\begin{document}
\author{
    \IEEEauthorblockN{
        Zhihao Li, Kun Li,  Boyang Ma,  Minghui Xu, Yue Zhang,  and Xiuzhen Cheng  
    \IEEEauthorblockA{
       \\Shandong University 
    }
}
}

\maketitle

\begin{abstract}
The Model Context Protocol (MCP) has emerged as a widely adopted mechanism for connecting large language models to external tools and resources. While MCP promises seamless extensibility and rich integrations, it also introduces a substantially expanded attack surface: any plugin can inherit broad system privileges with minimal isolation or oversight. In this work, we conduct the first large-scale empirical analysis of MCP security risks. We develop an automated static analysis framework and systematically examine 2{,}562 real-world MCP applications spanning 23 functional categories. Our measurements reveal that network and system resource APIs dominate usage patterns, affecting 1,438 and 1,237 servers respectively, while file and memory resources are less frequent but still significant. We find that Developer Tools and API Development plugins are the most API-intensive, and that less popular plugins often contain disproportionately high-risk operations. Through concrete case studies, we demonstrate how insufficient privilege separation enables privilege escalation, misinformation propagation, and data tampering. Based on these findings, we propose a detailed taxonomy of MCP resource access, quantify security-relevant API usage, and identify open challenges for building safer MCP ecosystems, including dynamic permission models and automated trust assessment. 
\end{abstract}


\section{Introduction}

Over the past year, the Model Context Protocol (MCP)~\cite{mcp-intro,mcp-spec} has rapidly emerged as a cornerstone for connecting large language models to the real world. By standardizing how AI systems discover, invoke, and coordinate external tools, MCP has fueled a wave of innovation—enabling everything from automated code generation to multi-modal content workflows. Yet while the protocol promises seamless extensibility and rich integrations, it also quietly redefines the attack surface of AI-driven applications. Underneath its clean JSON-RPC interface lies a powerful but underexplored question: \textit{What happens when thousands of independently developed plugins gain system-level access under the guise of helpful automation?} 
\looseness=-1

Unlike traditional mobile platforms that enforce runtime permission checks and sandboxing~\cite{DEMO,AutoCog,DBLP:conf/ccs/ZhangYXYGNWZ13,PScout,DBLP:conf/ccs/FeltCHSW11,DBLP:conf/ccs/BarreraKOS10,DBLP:conf/kbse/MalviyaTLXSJ23}, MCP servers typically run natively on local machines with few, if any, isolation boundaries. This architecture assumes high levels of implicit trust: any plugin installed by the user inherits powerful primitives, such as file I/O, process spawning, and unrestricted network communication. While these capabilities make MCP appealing to developers, they also raise significant security concerns, especially as the ecosystem grows rapidly without uniform security baselines.

To systematically measure this emerging risk, we developed an automated static analysis framework that scanned 2,562 real-world MCP applications spanning 23 functional categories. Our study produced several key findings. First, network and system resource APIs are by far the most prevalent, with 1,438 and 1,237 affected servers respectively, highlighting that most plugins rely heavily on privileged operations. By comparison, file resource threats were identified in 613 servers, and memory-related risks were relatively rare, with only 25 occurrences. We further observed that Developer Tools and API Development plugins accounted for the highest concentration of high-risk API calls, exceeding 500 calls per category. Interestingly, plugins in the lowest GitHub star range (0–10 stars) were responsible for 1,837 total API calls, significantly outpacing more popular or mature projects. This suggests that less widely reviewed plugins often integrate aggressive system interactions without appropriate safeguards.

Beyond aggregate statistics, we analyzed individual high-risk cases. For example, we demonstrated how a widely used blog-publisher plugin combines system-level \texttt{subprocess.run()} calls with unrestricted file copying, creating a clear Privilege Escalation Risk. Similarly, a \textit{twitter-mcp} integration exhibited Misinformation Risk by allowing silent manipulation of tweet content and embedded metadata. A web-research server presented a Data Tampering Risk, enabling attackers to inject biased content or exfiltrate user queries. Collectively, these examples illustrate the real-world consequences of insufficient privilege separation in MCP environments.

These findings highlight not only the urgent need for least-privilege design and systematic auditing in MCP development, but also a deeper challenge: how to reconcile the flexibility AI agents demand with the security guarantees users expect. We envision future directions such as dynamic permission models that respond to natural language intents in real time, platform-adaptive access controls, and automated certification pipelines to assess plugin trustworthiness before deployment. As MCP continues to evolve into the de facto interface between language models and the operating system, establishing robust foundations for security will be critical to realizing its promise without sacrificing safety.

Our contributions can be summarized as follows: 
\begin{itemize}
    \item   \textbf{Taxonomy of MCP Resource Access.}
We develop a detailed classification of the resource types exposed through MCP, systematically categorizing file, memory, network, and system resources as they relate to plugin functionality and potential security boundaries.

   \item   \textbf{Empirical Measurement of API Usage.}
We perform large-scale measurement of API invocation behaviors across 2,562 real-world MCP plugins, quantifying the prevalence of resource access patterns and analyzing associated security risks, including privilege escalation, data tampering, and misinformation threats.

   \item   \textbf{Identification of Open Research Questions.}
Based on our findings, we highlight a set of open questions and future directions such as dynamic permission models, scalable trust assessment, and automated policy enforcement—to guide the development of safer and more accountable MCP platforms.
\end{itemize}

\section{Background}

\subsection{MCP in a Nutshell}

The Model Context Protocol (MCP) is an open, JSON-RPC-based communication standard introduced by Anthropic in late 2024~\cite{mcp-intro}. It is designed to address a critical challenge in the evolution of AI systems: enabling large language models (LLMs) to interact seamlessly with external tools, APIs, and data sources in a unified, secure, and extensible way. Unlike previous ad hoc integrations where each tool or service required custom logic and manual data handling, MCP offers a general-purpose protocol for tool discovery, function invocation, and structured data exchange. As such, MCP plays a foundational role in advancing the capabilities of LLM-driven applications from passive language models to fully interactive and autonomous agents. \looseness=-1

\subsection{MCP Architecture}
\label{subsec:mcparch}

MCP employs a client-server architecture with modular components that enable separation of responsibilities, deployment extensibility, and interoperability between language models and external systems. \textit{MCP Hosts} are applications interfacing with large language models, such as Claude Desktop or Cursor IDE, responsible for initiating connections and providing user-facing runtime environments. \textit{MCP Clients} within each host maintain persistent connections with MCP Servers, orchestrating tool discovery, request formatting, and result delivery. \textit{MCP Servers} act as core capability providers, exposing structured APIs for tools (executable functions with defined schemas), resources (structured or unstructured data), and prompts (reusable instruction templates for guiding model behavior).

\begin{figure*}[htbp]
  \centering
  \includegraphics[width=\linewidth]{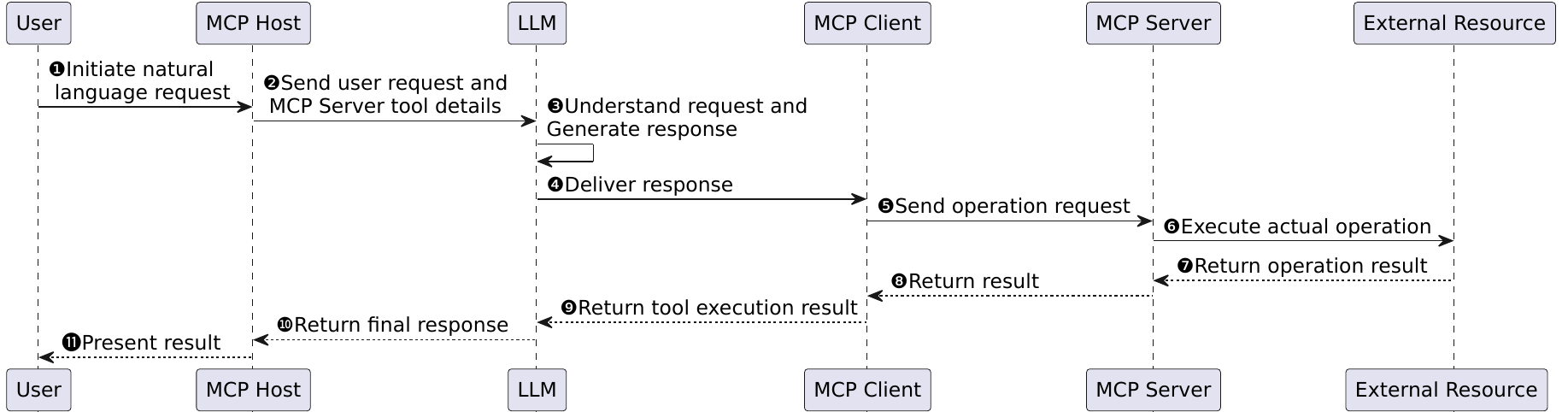}
  \caption{Workflow of MCP protocol.}
  \vspace{-5pt}
  \label{fig:mcp communication process}
\end{figure*}
    

\subsection{MCP Communication}

As shown in ~\autoref{fig:mcp communication process},  the workflow illustrates how a user’s natural language request is progressively interpreted, delegated to external tools, and resolved through coordinated interactions among the Host, Client, Server, and the LLM itself.
The interaction workflow begins when the user initiates a natural language request through the MCP Host interface, expressing an intent that may require access to external tools or resources (Step \ding{182}). Upon receiving this request, the MCP Host transmits both the raw user input and detailed metadata describing available MCP Server tools to the LLM component, ensuring that the model has full visibility into the capabilities it can potentially leverage (Step \ding{183}). The LLM then analyzes the request to interpret the user’s objective, performs contextual reasoning, and determines whether invoking an external tool is necessary to fulfill the task (Step \ding{184}). If so, it generates a structured invocation request specifying the tool to be called and any required parameters, and delivers this request back to the MCP Client for further processing (Step \ding{185}). The MCP Client takes responsibility for constructing a formal operation request compliant with the MCP Server interface and dispatches this request to the server component (Step \ding{186}). Upon receiving the operation request, the MCP Server executes the actual action, which may involve interacting with external environments, querying files, accessing system resources, or invoking third-party APIs, depending on the plugin’s implementation (Step \ding{187}). After the execution is completed (Step \ding{188}), the server packages the operation result and returns it to the MCP Client, preserving all relevant context and output data (Step \ding{189}). The MCP Client relays this result back to the LLM, which integrates the returned data with its internal reasoning context to generate a coherent, human-readable response (Step \ding{190}). Finally, the completed response is transmitted to the MCP Host (Step \ding{191}), which presents the information to the user, concluding the interaction cycle (Step \blackcircled{11}).

\section{Taxonomy of MCP-Accessible Resources}

Through our manual analysis of MCP plugin codebases, we observed that MCP Servers frequently interact with four principal categories of system resources through APIs: File, Memory, Network, and System Resources. Each category reflects a different dimension of operational capability and introduces distinct security implications.

\begin{itemize}
\item \textbf{File Resources} represent one of the most commonly accessed categories. Plugins often require reading or writing files to support legitimate functionality, such as document indexing, log processing, configuration management, or code generation workflows. Typical operations include reading project files, writing output artifacts, or enumerating directories.

\item \textbf{Memory Resources} are accessible through APIs that enable inter-process communication, shared memory allocation, or memory inspection. MCPs may leverage these capabilities for tasks such as performance monitoring, data exchange between processes, or advanced debugging.

\item \textbf{Network Resources} are essential for many MCPs, supporting features like remote data retrieval, web service integration, and notification delivery. Common examples include sending HTTP requests, establishing persistent connections, or performing DNS lookups to locate external endpoints.

\item \textbf{System Resources} encompass interfaces and APIs that allow low-level control over process management, environment configuration, and hardware interaction. Typical uses include spawning subprocesses to run helper tools, modifying environment variables to adjust runtime behavior, or accessing system configuration data.

\end{itemize}

\section{Motivation and Research Question}

As discussed in \S\ref{subsec:mcparch}, the MCP architecture comprises three components: the Host (running the AI application), the Client (which communicates with the Server), and the Server (the backend plugin handler). Unlike mobile platforms such as Android or iOS, which enforce runtime permission checks and app isolation~\cite{DEMO,AutoCog,DBLP:conf/ccs/ZhangYXYGNWZ13,PScout,DBLP:conf/ccs/FeltCHSW11,DBLP:conf/ccs/BarreraKOS10,DBLP:conf/kbse/MalviyaTLXSJ23,DBLP:conf/kbse/MalviyaLKTSJ22,DBLP:conf/kbse/HolavanalliMNRSKZ13}, MCP Servers often run locally with high privileges and inherently trust third-party code, creating opportunities for abuse by malicious or poorly implemented plugins. This architecture assumes implicit trust without built-in mechanisms to constrain what plugins can access or perform. As a result, even seemingly benign plugins can freely invoke sensitive APIs to read arbitrary files, initiate outbound network connections, monitor memory, or interact with system processes without user awareness or consent. These risks are not hypothetical. For example, a plugin claiming to help search local documents could exfiltrate SSH keys, while a Git workflow plugin could silently install backdoors. To address this growing attack surface, we present a static analysis framework that systematically examines MCP plugins through the lens of system-level API usage. Instead of relying on protocol inspection or runtime monitoring, which can miss implicit behaviors or be bypassed, our approach focuses on how plugins are written, what system APIs they invoke, how these calls propagate through helper functions, and whether they match known security-sensitive patterns. We argue that static API analysis offers a language-agnostic, scalable, and explainable foundation for auditing plugin behavior in an open ecosystem with minimal security vetting and widely varying developer practices.

Despite the flexibility and openness of the MCP plugin ecosystem, its architecture raises critical security questions that remain underexplored. To bridge this gap, we center our study around three research questions designed to illuminate the nature and scale of MCP security threats.

\begin{itemize}[left=0.7cm]
    \item [\textbf{RQ1:}] \textit{What are the main security threat types in MCP Servers, and how are they distributed in practice?}

    \item [\textbf{RQ2:}] \textit{How does the adoption of system APIs vary across different MCP application categories?}

\item [\textbf{RQ3:}] \textit{What is the correlation between MCP application popularity and code containing dangerous APIs?}

\end{itemize}

\section{Design and Implementation}

\subsection{Design}
\label{subsec:design}

The framework operates in three distinct phases, each addressing a critical aspect of large-scale, multi-language plugin assessment. The first phase focuses on collecting and standardizing diverse codebases to build a consistent analysis corpus, while the second phase performs fine-grained API usage analysis to identify operations related to sensitive system resources. By combining automated crawling, multi-language parsing, and targeted API classification, the framework provides a scalable foundation for quantifying risk across a wide spectrum of MCP plugin implementations.

\noindent\textbf{(P-I): Code Collection and Preprocessing.} 
The analysis of MCP plugins poses significant challenges due to their large scale and cross-language implementations. We address these challenges through automated source code collection and standardized preprocessing. Our approach employs web crawling to extract plugin entries from popular platforms such as MCP Market, subsequently downloading source code by parsing the associated repository links. Since repositories exhibit diverse directory structures, file types, and encoding formats, we implement standardized preprocessing to normalize source code structure, standardize encodings, and remove temporary files and build artifacts, creating a consistent dataset for security analysis.

Beyond the source code, the system synchronously collects key metadata for each plugin, such as repository category, primary language, popularity (e.g., GitHub's star count), and platform tags. By integrating external data sources, a multi-level mapping of “Server–Repository–Category–Attribute” is constructed, laying the foundation for subsequent threat distribution analysis and correlation modeling. Additionally, the system automatically performs deduplication and validation for duplicate or invalid repositories to ensure the uniqueness and representativeness of analysis targets.


\vspace{2mm}
\noindent\textbf{(P-III):  Multi-Dimensional API Analysis.} 
This phase performs comprehensive API analysis to identify security-critical operations within the normalized source code. We focus on detecting high-risk API invocation behaviors at the source code level, leveraging multi-language parsing techniques and threat signature databases to achieve accurate identification of resource abuse and potential security risks. Based on MCP API types and their associated resources, we classify four categories of resource-related critical APIs as primary analysis targets: file, system, network, and memory. The typical operations and potential risk points for these four API categories are presented in Table~\ref{tab:api}.


\begin{table}[!ht]
\caption{API Categories and Security Risk Examples}
\centering
\scriptsize
\begin{tabular}{cll}
\toprule[1.5pt]
\textbf{Resource} & \textbf{API/Operation Examples} & \textbf{Potential Security Risks} \\
\midrule

File & \makecell[l]{\texttt{open()}\\ \texttt{read()} \\\texttt{write()}
}& \makecell[l]{Unauthorized file access, \\hardcoded paths, \\TOCTOU race conditions} \\
\midrule

System & \makecell[l]{\texttt{os.system()}\\ \texttt{subprocess.call()}\\ \texttt{fork()}\\ \texttt{exec()}} & \makecell[l]{Command injection (RCE), \\improper process management, \\privilege escalation} \\
\midrule

Network & \makecell[l]{\texttt{socket.bind}\\ \texttt{connect()}\\ \texttt{dns.resolver.query()}} & \makecell[l]{Open high-risk ports, \\unencrypted communications, \\DNS hijacking} \\
\midrule

Memory & \makecell[l]{\texttt{strcpy()}\\ \texttt{malloc()}\\ \texttt{ctypes.CDLL()}\\ \texttt{create\_string\_buffer()}} & \makecell[l]{Buffer overflow, \\malicious library injection, \\fixed buffer overflow} \\

\bottomrule[1.5pt]
\end{tabular}
\label{tab:api}
\end{table}

\vspace{2mm}
\noindent\textbf{(P-III): Result Aggregation and Reporting.} 
In the last phase, the framework consolidates multi-dimensional API risk detection results and transforms them into actionable insights through categorization and visualization, enabling comprehensive threat profiling from individual plugins to platform-wide analysis. 

\subsection{Implementation}
\label{subsec:implementation}

The framework is implemented as a modular pipeline combining automated crawling, multi-language parsing, and high-performance batch processing. For code collection, we developed custom crawler scripts that traverse MCP Market indexes, extract repository URLs, and recursively download source code and associated metadata. Preprocessing scripts apply directory pruning rules to remove non-essential content (e.g., \texttt{node\_modules}, \texttt{venv}, binaries) and normalize all files to UTF-8 encoding and consistent line endings. For program analysis, the system integrates language-specific AST parsers (Python AST, JavaScript ESTree, JavaParser) with fallback regex-based scanning to handle syntax errors and edge cases. A curated API signature database is embedded in the analysis engine, enabling precise matching of invocation patterns and parameter sensitivities. The framework supports parallel scanning across multiple repositories, automatically skips third-party dependencies, and archives per-run results in each repository’s \texttt{.git} directory for incremental comparison. For reporting, it generates Markdown outputs and visual charts using structured JSON exports, facilitating integration with external dashboards or security tools.

\section{Evaluation}

\subsection{Experiment Setup}
To validate the proposed approach, we collected 2,562 MCP applications from MCP Market (url: mcpmarket.com), an MCP server aggregation and distribution platform. MCP Market gathers globally accessible MCP Server tools, offers documentation, community support, and provides development templates and debugging tools for rapid MCP toolchain development. We also recorded features such as category and GitHub star count for correlation analysis.

\subsection{Experiment Results}

\begin{figure}[!tb]
  \centering
  \includegraphics[width=0.95\columnwidth]{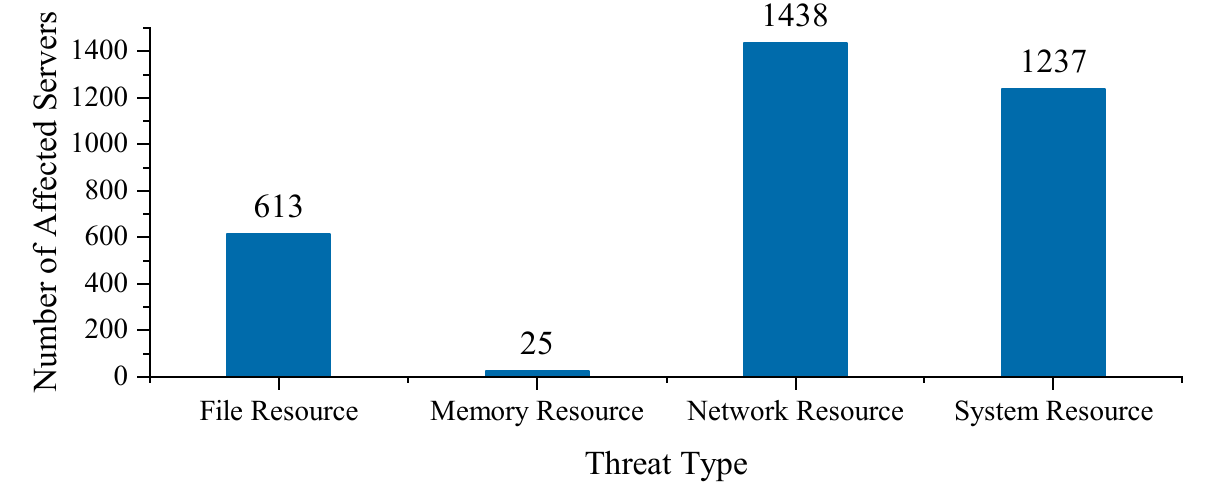}
  \vspace{-5pt}
  \caption{Distribution of MCP Server threat types.}
  \label{fig:threat-dist}
\end{figure}

\vspace{2mm}
\noindent\textbf{Answer to RQ1}: We conducted static analysis on the collected 2,562 MCP applications using our proposed framework. ~\autoref{fig:threat-dist} presents the number of MCP servers affected by each threat type. The figure was generated using a standard bar chart visualization, where each bar represents the total count of servers invoking APIs associated with a specific resource category. For clarity, numeric labels were added above each bar to indicate exact counts, and the axes were annotated to show the distribution of affected servers across threat types.
As illustrated in Figure~\ref{fig:threat-dist}, threats originating from network and system resources significantly outnumber those from other resource categories, affecting 1,438 and 1,237 servers respectively.   Furthermore, 613 servers exhibit file resource threats, while only 25 servers are affected by memory resource threats. \looseness=-1 

\vspace{1.5mm}
 \begin{mdframed}[backgroundcolor=green!2] 
\noindent\textit{
System and network resource threats are by far the most prevalent across MCP servers, indicating widespread exposure to high-impact attack surfaces such as remote code execution and unauthorized network communication. In contrast, file and memory resource threats are less common in absolute numbers but remain important due to their potential for sensitive data compromise and privilege escalation.
}
\end{mdframed}
\vspace{1.5mm}

\vspace{2mm}
\noindent\textbf{Answer to RQ2}:
We categorized MCP applications according to their API resource usage patterns and examined the distribution across four primary resource types: file resource, memory resource, network resource, and system. The results are summarized in Table~\ref{tab:distribution}, which reports the count of API calls per resource type within each application category. The counts in the table were derived by scanning each MCP application’s normalized source code and tallying the number of detected API calls per resource type. Each row aggregates the totals for all plugins belonging to the corresponding application category.  As shown in  \autoref{tab:distribution},   the distribution of API calls across different MCP application categories and resource types. Overall, Developer Tools have the highest total number of API calls (626), followed by API Development (511) and Data Science \& ML (196). In contrast, categories such as Featured, E-commerce Solutions, and Mobile Development have relatively few API calls, each with fewer than 15 entries.

\vspace{1.5mm}
 \begin{mdframed}[backgroundcolor=green!2] 
\noindent\textit{The security insight is that categories with high API usage (especially Developer Tools and API Development) pose a higher risk due to their broader resource access, making them priority targets for thorough security assessment.
}
\end{mdframed}
\vspace{1.5mm}



\begin{table} 
\centering
\scriptsize
\caption{API Call Distribution by MCP Application Category}
\setlength\tabcolsep{2.3pt}
\begin{tabular}{lccccc}
\toprule[1.5pt]
\makecell{\textbf{Dimension}} & 
\makecell{\textbf{Total}} & 
\makecell{\textbf{File}\\ \textbf{Resource}} & 
\makecell{\textbf{Memory}\\ \textbf{Resource}} & 
\makecell{\textbf{Network}\\ \textbf{Resource}} & 
\makecell{\textbf{System}\\ \textbf{Resource}} \\
\midrule
API Development                & 511 & 77  & 4  & 313 & 195 \\
Analytics \& Monitoring        & 46  & 9   & 0  & 36  & 15  \\
Browser Automation             & 28  & 5   & 0  & 12  & 21  \\
Cloud Infrastructure           & 49  & 11  & 0  & 27  & 31  \\
Collaboration Tools            & 72  & 12  & 2  & 64  & 29  \\
Content Management             & 13  & 3   & 0  & 11  & 3   \\
Data Science \& ML             & 196 & 54  & 1  & 131 & 100 \\
Database Management            & 42  & 9   & 2  & 22  & 26  \\
Deployment \& DevOps           & 67  & 21  & 0  & 30  & 48  \\
Design Tools                   & 28  & 12  & 0  & 21  & 16  \\
Developer Tools                & 626 & 156 & 7  & 325 & 336 \\
E-commerce Solutions           & 13  & 0   & 0  & 12  & 4   \\
Featured                       & 5   & 1   & 0  & 3   & 3   \\
Game Development               & 16  & 5   & 0  & 12  & 11  \\
Learning \& Documentation      & 52  & 14  & 1  & 35  & 22  \\
Marketing Automation           & 12  & 1   & 0  & 10  & 5   \\
Mobile Development             & 11  & 6   & 0  & 3   & 9   \\
Official                       & 5   & 2   & 0  & 4   & 2   \\
Other                          & 497 & 151 & 2  & 229 & 235 \\
Productivity \& Workflow       & 147 & 52  & 6  & 82  & 102 \\
Security \& Testing            & 77  & 24  & 4  & 34  & 62  \\
Social Media Management        & 22  & 1   & 0  & 19  & 8   \\
Web Scraping \& Data Collection& 82  & 11  & 0  & 64  & 34  \\
\bottomrule[1.5pt]
\end{tabular}
\label{tab:distribution}
\end{table}

\vspace{2mm}
\noindent\textbf{Answer to RQ3}:
We categorized MCP applications based on their GitHub popularity, measured by star count, and examined the distribution of API calls across four primary resource types: file resource, memory resource, network resource, and system. The results are summarized in ~\autoref{tab:StarRange}, which reports the count of API calls per resource type within each star range interval. The counts were derived by scanning each plugin’s normalized source code and tallying the number of detected API calls corresponding to each resource category. Each row aggregates totals for all plugins falling within the respective GitHub star range. As shown in \autoref{tab:StarRange}, plugins with 0–10 stars account for the majority of API calls overall (1,837), while projects with higher star counts contribute progressively fewer calls, with the 50,000+ star range containing only a single plugin. \looseness=-1

\begin{table} 
\centering
\scriptsize
\caption{API Call Distribution by GitHub Star Range}
\setlength\tabcolsep{6pt}
\begin{tabular}{lrrrrr}
\toprule[1.5pt]
\makecell{\textbf{Star Range}} & 
\makecell{\textbf{Total}} & 
\makecell{\textbf{File}\\ \textbf{Resource}} & 
\makecell{\textbf{Memory}\\ \textbf{Resource}} & 
\makecell{\textbf{Network}\\ \textbf{Resource}} & 
\makecell{\textbf{System}\\ \textbf{Resource}} \\
\midrule
0--10         & 1837 & 367 & 10 & 1079 & 834 \\
11--100       & 504  & 122 & 7  & 240  & 238 \\
101--1000     & 164  & 86  & 4  & 77   & 112 \\
1001--10000   & 47   & 29  & 2  & 32   & 43  \\
10001--50000  & 9    & 8   & 2  & 9    & 9   \\
50000+        & 1    & 1   & 0  & 1    & 1   \\
\bottomrule[1.5pt]
\end{tabular}
\label{tab:StarRange}
\end{table}



\vspace{1.5mm}
 \begin{mdframed}[backgroundcolor=green!2] 
\noindent\textit{The data reveals that less popular MCP applications tend to implement more aggressive functionalities requiring dangerous APIs, while popular applications adopt conservative approaches. This suggests the community favors stable, secure applications over feature-rich but potentially risky ones. Popular applications appear to prioritize safety and reliability, avoiding extensive use of system-level, network, or memory manipulation APIs that could pose security risks.
}
\end{mdframed}
\vspace{1.5mm}

\section{Practical Security Threat Analysis}

To understand the practical impact of MCP security threats, we conducted detailed security risk (\textbf{R}) assessments on three mainstream MCP servers: \textit{blog-publisher}, \textit{twitter-mcp}, and \textit{web-research}, each representing different application domains. 


\noindent\textbf{(R1) Privilege Escalation Risk for \textit{blog-publisher.}}
This server manages blog content by executing local git commands via system calls. Our analysis revealed critical vulnerabilities arising from unsanitized input handling, enabling malicious operators to inject arbitrary commands during repository operations. Attackers could access entire git repositories, exposing source code, configurations, and credentials. The server bypasses GitHub's access controls through direct repository modification, facilitating unauthorized code injection. Combined system-level access (\texttt{subprocess.run()}) and unrestricted file operations (\texttt{shutil.copy()}) create substantial attack vectors for privilege escalation and data exfiltration.


\noindent\textbf{(R2) Misinformation Risk for \textit{twitter-mcp.}} This server provides comprehensive X (formerly Twitter) integration, enabling tweet posting, searching, and account management. Our security assessment uncovered significant potential for social engineering attacks and misinformation campaigns. Malicious operators could manipulate tweet content before publication, inserting propaganda, phishing links, or misleading information without user awareness. The server's network capabilities (\texttt{post}) combined with image processing functions (\texttt{Image.open}) enable sophisticated attacks, including malicious metadata embedding or fraudulent website redirection. Attackers could harvest user engagement data, monitor follower patterns, or manipulate trending topics through coordinated compromised accounts, potentially influencing public opinion or spreading misinformation at scale.

\noindent\textbf{(R3) Data Tampering Risk  \textit{web-research.}} This web search and research server presents unique risks related to information integrity and user privacy. Our analysis demonstrated how attackers could manipulate search results to inject biased content, phishing links, or malicious websites into research outputs. The server's integration with OpenAI services (\texttt{OPENAI()}) and external APIs creates opportunities for data interception and manipulation. Malicious operators could log all user queries, building detailed profiles of research interests and potentially sensitive information needs. Additionally, the server's environment configuration loading (\texttt{load\_dotenv()}) could expose API keys and credentials if improperly secured. The combination of network access (\texttt{post()}) and file operations enables attackers to exfiltrate research data or inject false information into academic or business research workflows, potentially leading to flawed decision-making based on compromised data.

\section{Open Questions and Future Directions}

Based on our comprehensive analysis of the MCP ecosystem, we propose three critical open questions: 
\begin{itemize}
    \item \textbf{Towards a Privilege Management Framework for MCP.} Our analysis reveals that MCP applications frequently operate with excessive privileges, accessing sensitive system resources without proper justification or user awareness. This underscores the urgent need for least-privilege API usage, comprehensive security auditing, and robust access control frameworks as foundational requirements for MCP development. However, this raises a fundamental question: How can we design an effective privilege management system that balances the flexibility MCP applications require with the security constraints users deserve? Unlike mobile platforms where apps have well-defined use cases~\cite{DBLP:conf/kbse/MalviyaTLXSJ23,DBLP:conf/kbse/MalviyaLKTSJ22,DBLP:conf/kbse/HolavanalliMNRSKZ13,DBLP:conf/kbse/BartelKTM12}, MCP applications serve as general-purpose intermediaries between AI agents and system resources, making it difficult to predetermine appropriate privilege boundaries.
    \item 
\textbf{Platform-Specific Security Adaptations.} Our domain-specific risk profiles suggest that security frameworks should incorporate context-aware controls tailored to specific business scenarios. However, desktop operating systems present unique challenges—Windows' capability-based security differs from Unix-like discretionary access controls, while macOS introduces sandboxing APIs. Should MCP security frameworks adapt to each platform's native security primitives, or would a unified cross-platform approach be more practical? Our analysis suggests leveraging platform-specific mechanisms, Windows' UAC, macOS's sandboxing, and Linux's namespace isolation. 
\item \textbf{Dynamic Permission Models for AI Agents.} Traditional permission systems rely on static declarations, but our findings show that MCP applications exhibit dynamic behavior based on AI agent requests. Can we develop context-aware permission systems that grant privileges just-in-time based on actual usage patterns rather than broad upfront declarations? This requires understanding the semantic relationship between natural language commands and required system resources—a challenge at the intersection of NLP and systems security that could fundamentally transform how we approach AI agent security.

\end{itemize}

\section{Related Work}

As an emerging protocol introduced by Anthropic in late 2024, MCP has attracted limited research attention. Ray et al.\cite{ray2025survey} survey MCP's architectural design and applications, identifying challenges like transmission latency and privilege escalation risks. Hou et al.\cite{hou2025modelcontextprotocolmcp} analyze MCP's lifecycle and security threats including \textit{tool name conflicts}, \textit{sandbox escapes}, and \textit{configuration drift}. 
Current MCP security research focuses on dynamic defenses. Kumar et al.\cite{kumar2025mcp} propose MCP Guardian, enforcing token-based authentication and WAF rules to block command injection attacks. Narajala et al.\cite{narajala2025enterprise} extend this with enterprise frameworks incorporating behavioral analysis. These approaches provide reactive monitoring but cannot prevent malicious code deployment or detect pre-execution risks.
Research on static security analysis of MCP Server source code remains virtually unexplored. We propose systematic API resource classification and static analysis that categorizes MCP server operations into four resource types--file, network, system, and memory--enabling comprehensive threat profiling and lightweight pre-deployment security evaluation across the MCP ecosystem.

\section{Conclusion}

In this paper, we conducted the first large-scale study of MCP plugin security, analyzing 2,562 applications to characterize their resource access patterns and associated risks. Our findings show that system and network APIs are pervasive, exposing substantial attack surfaces for privilege escalation and data manipulation. By introducing a detailed taxonomy, quantitative measurements, and real-world case studies, we highlight the urgent need for stronger isolation and policy enforcement. We hope these insights will guide future efforts to develop safer, more accountable MCP ecosystems.

\bibliographystyle{IEEEtran}
\bibliography{references}

\end{document}